\documentclass[aps,prl,showpacs,twocolumn,superscriptaddress,letterpaper,longbibliography]{revtex4-1}
\usepackage{graphicx}
\usepackage{dcolumn}
\usepackage{bm}
\usepackage{amsmath, amssymb}%
\usepackage{epstopdf}
\usepackage{color}
\usepackage[normalem]{ulem}
\usepackage{multirow}
\usepackage{enumitem} 
\usepackage{csquotes}
\usepackage{makecell}
\usepackage{hyperref}  
\hypersetup{breaklinks = true, colorlinks = true, citecolor = blue, linkcolor = blue, urlcolor = blue}
\usepackage{cleveref}
\usepackage{float}

\usepackage[
singlelinecheck=false
]{caption}
\usepackage[font=small,justification=raggedright, format=plain]{caption}

\floatstyle{plaintop}
\restylefloat{table}

\newcommand{\neumu}{$\nu_{\mu}$}

\newcommand{\uB}{MicroBooNE}

\newcommand{\CCIpOpi}{CC1p0$\pi$}

\newcommand{\CosThetaMu}{$\cos\theta_{\mu}$}

\newcommand{\argon}{$^{40}$Ar\,\,}

\begin{document}

\title{First Measurement of Differential Charged Current Quasielastic--like $\nu_{\mu}$--Argon Scattering Cross Sections with the MicroBooNE Detector}
\date{\today}

 
 \renewcommand{\arraystretch}{1.3}


\newcommand{\Bern}{Universit{\"a}t Bern, Bern CH-3012, Switzerland}
\newcommand{\BNL}{Brookhaven National Laboratory (BNL), Upton, NY, 11973, USA}
\newcommand{\UCSB}{University of California, Santa Barbara, CA, 93106, USA}
\newcommand{\Cambridge}{University of Cambridge, Cambridge CB3 0HE, United Kingdom}
\newcommand{\StKates}{St. Catherine University, Saint Paul, MN 55105, USA}
\newcommand{\Chicago}{University of Chicago, Chicago, IL, 60637, USA}
\newcommand{\Cincinnati}{University of Cincinnati, Cincinnati, OH, 45221, USA}
\newcommand{\CSU}{Colorado State University, Fort Collins, CO, 80523, USA}
\newcommand{\Columbia}{Columbia University, New York, NY, 10027, USA}
\newcommand{\Davidson}{Davidson College, Davidson, NC, 28035, USA}
\newcommand{\FNAL}{Fermi National Accelerator Laboratory (FNAL), Batavia, IL 60510, USA}
\newcommand{\Granada}{Universidad de Granada, E-18071, Granada, Spain}
\newcommand{\Harvard}{Harvard University, Cambridge, MA 02138, USA}
\newcommand{\IIT}{Illinois Institute of Technology (IIT), Chicago, IL 60616, USA}
\newcommand{\KSU}{Kansas State University (KSU), Manhattan, KS, 66506, USA}
\newcommand{\Lancaster}{Lancaster University, Lancaster LA1 4YW, United Kingdom}
\newcommand{\LANL}{Los Alamos National Laboratory (LANL), Los Alamos, NM, 87545, USA}
\newcommand{\Manchester}{The University of Manchester, Manchester M13 9PL, United Kingdom}
\newcommand{\MIT}{Massachusetts Institute of Technology (MIT), Cambridge, MA, 02139, USA}
\newcommand{\Michigan}{University of Michigan, Ann Arbor, MI, 48109, USA}
\newcommand{\Minnesota}{University of Minnesota, Minneapolis, MN, 55455, USA}
\newcommand{\NMSU}{New Mexico State University (NMSU), Las Cruces, NM, 88003, USA}
\newcommand{\Otterbein}{Otterbein University, Westerville, OH, 43081, USA}
\newcommand{\Oxford}{University of Oxford, Oxford OX1 3RH, United Kingdom}
\newcommand{\PNNL}{Pacific Northwest National Laboratory (PNNL), Richland, WA, 99352, USA}
\newcommand{\Pitt}{University of Pittsburgh, Pittsburgh, PA, 15260, USA}
\newcommand{\Rutgers}{Rutgers University, Piscataway, NJ, 08854, USA}
\newcommand{\StMarys}{Saint Mary's University of Minnesota, Winona, MN, 55987, USA}
\newcommand{\SLAC}{SLAC National Accelerator Laboratory, Menlo Park, CA, 94025, USA}
\newcommand{\SDSMT}{South Dakota School of Mines and Technology (SDSMT), Rapid City, SD, 57701, USA}
\newcommand{\Syracuse}{Syracuse University, Syracuse, NY, 13244, USA}
\newcommand{\TelAviv}{Tel Aviv University, Tel Aviv, Israel, 69978}
\newcommand{\Tennessee}{University of Tennessee, Knoxville, TN, 37996, USA}
\newcommand{\UTA}{University of Texas, Arlington, TX, 76019, USA}
\newcommand{\Tufts}{Tufts University, Medford, MA, 02155, USA}
\newcommand{\VTech}{Center for Neutrino Physics, Virginia Tech, Blacksburg, VA, 24061, USA}
\newcommand{\Warwick}{University of Warwick, Coventry CV4 7AL, United Kingdom}
\newcommand{\Yale}{Wright Laboratory, Department of Physics, Yale University, New Haven, CT, 06520, USA}

\affiliation{\Bern}
\affiliation{\BNL}
\affiliation{\UCSB}
\affiliation{\Cambridge}
\affiliation{\StKates}
\affiliation{\Chicago}
\affiliation{\Cincinnati}
\affiliation{\CSU}
\affiliation{\Columbia}
\affiliation{\Davidson}
\affiliation{\FNAL}
\affiliation{\Granada}
\affiliation{\Harvard}
\affiliation{\IIT}
\affiliation{\KSU}
\affiliation{\Lancaster}
\affiliation{\LANL}
\affiliation{\Manchester}
\affiliation{\MIT}
\affiliation{\Michigan}
\affiliation{\Minnesota}
\affiliation{\NMSU}
\affiliation{\Otterbein}
\affiliation{\Oxford}
\affiliation{\PNNL}
\affiliation{\Pitt}
\affiliation{\Rutgers}
\affiliation{\StMarys}
\affiliation{\SLAC}
\affiliation{\SDSMT}
\affiliation{\Syracuse}
\affiliation{\TelAviv}
\affiliation{\Tennessee}
\affiliation{\UTA}
\affiliation{\Tufts}
\affiliation{\VTech}
\affiliation{\Warwick}
\affiliation{\Yale}

\author{P.~Abratenko} \affiliation{\Tufts} 
\author{M.~Alrashed} \affiliation{\KSU}
\author{R.~An} \affiliation{\IIT}
\author{J.~Anthony} \affiliation{\Cambridge}
\author{J.~Asaadi} \affiliation{\UTA}
\author{A.~Ashkenazi} \affiliation{\MIT}
\author{S.~Balasubramanian} \affiliation{\Yale}
\author{B.~Baller} \affiliation{\FNAL}
\author{C.~Barnes} \affiliation{\Michigan}
\author{G.~Barr} \affiliation{\Oxford}
% ...

\collaboration{The MicroBooNE Collaboration} \thanks{microboone\_info@fnal.gov}\noaffiliation


\begin{abstract}
We report on the first measurement of flux-integrated single differential cross sections for charged-current (CC) muon neutrino  ($\nu_{\mu}$) scattering on argon with a muon and a 
proton in the final state, \argon$(\nu_{\mu},\mu p)X$. 
The measurement was carried out using the Booster Neutrino Beam at Fermi National Accelerator Laboratory and the MicroBooNE liquid argon time projection chamber detector with 
an exposure of $4.59 \times 10^{19}$ protons on target. 
Events are selected to enhance the contribution of CC quasielastic (CCQE) interactions.
The data are reported in terms of a 
total cross section as well as single differential cross sections in final state muon and proton kinematics. 
We measure the integrated per-nucleus CCQE--like cross section (i.e. for interactions leading to a muon, one proton and no pions above detection threshold) of $(4.93\pm 0.76_\text{stat} \pm 1.29_\text{sys} ) \times 10^{-38} \textrm{cm}^2$, in good agreement with theoretical calculations. 
The single differential cross sections are also 
in overall good agreement with theoretical predictions, except at very forward muon scattering angles that correspond to low momentum-transfer events.
\end{abstract}

\maketitle


Measurements of neutrino oscillation serve as a valuable tool for extracting 
neutrino mixing angles, mass-squared differences, and the CP violating phase, 
as well as for searching for new physics beyond the standard model in the electroweak sector~\cite{pdg2018,T2KNature20}.

Neutrinos oscillate as a function of their propagation distance divided by their energy.
In accelerator-based oscillation experiments, the neutrino propagation distance is well defined.
However, as these experiments do not use mono-energetic neutrino beams 
~\cite{AguilarArevalo:2008yp,Aliaga:2016av,Abe:2012av},
the accuracy to which they can extract neutrino oscillation parameters depends 
on their ability to determine the individual energy of the detected neutrinos.
This requires detailed understanding of the fundamental interactions 
of neutrinos with atomic nuclei that comprise neutrino detectors.

Understanding the interaction of neutrinos with argon nuclei is of particular importance
as a growing number of neutrino oscillation experiments employ liquid argon time projector chamber (LArTPC) neutrino detectors.
These include the Deep Underground Neutrino Experiment (DUNE)~\cite{Abi:2020wmh,Abi:2020evt,Abi:2020oxb,Abi:2020loh}, 
which aims to measure the neutrino CP-violating phase and mass hierarchy, 
and the Short Baseline Neutrino (SBN) program~\cite{Antonello:2015lea}, 
that is searching for physics beyond the Pontecorvo--Maki--Nakagaw--Sakata (PMNS) matrix model of neutrino mixing.

Experimentally, the energy of interacting neutrinos is determined from the measured momenta of particles 
that are emitted following the neutrino interaction in the detector.
Many accelerator-based oscillation studies focus on measurements of 
charged-current (CC) neutrino-nucleon quasielastic (QE) scattering 
interactions~\cite{Anderson:2012jds,Nakajima:2010fp,Aguilar-Arevalo:2013dva,Abe:2014iza,Carneiro:2019jds,Abratenko:2019jqo,Fiorentini:2013ezn,Betancourt:2017uso,Walton:2014esl,Abe:2018pwo}, 
where the neutrino removes a single intact nucleon from the nucleus without producing any additional particles.
This choice is guided by the fact that CCQE reactions can be reasonably well approximated as two-body interactions, 
and their experimental signature of a correlated muon-proton pair is relatively straightforward to measure.
Therefore, precise measurements of CCQE processes are expected to allow precise reconstruction 
of neutrino energies with discovery-level accuracy~\cite{Mosel:2013fxa}. 

A working definition for identifying CCQE interactions in experimental measurements requires the identification of a 
neutrino interaction vertex with an outgoing lepton, exactly one outgoing proton, and no additional particles;
We refer to these herein as “CCQE--like” events.
This definition can include contributions from non--CCQE interactions that lead to the production
of additional particles that are absent from the final state due to nuclear effects such as pion absorption or have momenta that are below the experimental detection threshold.

Existing data on neutrino CCQE--like interactions come from experiments using various energies and target nuclei~\cite{Formaggio:2013kya}. These primarily include measurements 
of CCQE--like muon neutrino ($\nu_\mu$) cross sections for interactions where a muon and no pions were detected,
with~\cite{Fiorentini:2013ezn,Betancourt:2017uso,Walton:2014esl,Abe:2018pwo} 
and without~\cite{Anderson:2012jds,Nakajima:2010fp,Aguilar-Arevalo:2013dva,Abe:2014iza,Carneiro:2019jds,Abratenko:2019jqo} 
requiring the additional detection of a proton in the final state.
While most relevant for LArTPC based oscillation experiments,
no measurements of CCQE--like cross sections on \argon\ with the detection of a proton in the final state exist.

This letter presents the first measurement of exclusive CCQE--like neutrino-argon interaction cross-sections,
measured using the MicroBooNE liquid argon time projection chamber (LArTPC).
Our data serve as the first study of exclusive CCQE--like differential cross sections on \argon 
as well as a benchmark for theoretical models of $\nu_{\mu}$-\argon interactions, which are  
key for performing a precise extraction of oscillation parameters by future LArTPC oscillation experiments.

We focus on a specific subset of CCQE--like interactions, denoted here as CC1p0$\pi$, where the contribution of CCQE interactions is enhanced~\cite{Adams:2018lzd}.
These include charged-current \neumu-$^{40}$Ar scattering events with a detected muon and exactly one proton, with momenta greater than 100 MeV/$c$ and 300 MeV/$c$, respectively. 
The measured muon-proton pairs are required to be co-planar with small missing transverse momentum and minimal residual activity near the interaction vertex that is not associated 
with the measured muon or proton.
For these CC1p0$\pi$ events we measure the flux--integrated $\nu_{\mu}$-\argon total and differential cross sections in muon and proton  momentum and angle, and as a function of the calorimetric measured energy and the reconstructed momentum transfer.

The measurement uses data from the \uB\  LArTPC detector~\cite{Acciarri:2016smi}, which is the first of a
series of LArTPCs to be used for precision oscillation measurements~\cite{Antonello:2015lea,Tortorici:2018yns,Abi:2020wmh,Abi:2020evt,Abi:2020oxb,Abi:2020loh}.
	The \uB\ detector has an active mass of 85 tons 
	and is located along the Booster Neutrino Beam (BNB) at Fermilab, 463 m downstream from the target.
	The BNB energy spectrum extends to 2 GeV and peaks around 0.7 GeV~\cite{AguilarArevalo:2008yp}.

    A neutrino is detected by its interaction with an argon nucleus in the LArTPC.
	The secondary charged particles produced in the interaction travel through the liquid argon,
	leaving a trail of ionization electrons that drift horizontally  and  transverse  to  the  neutrino beam direction in an electric field of 273 V/cm, to  
	a system of three anode wire planes located 2.5 m from the cathode plane.
	The Pandora tracking package~\cite{Acciarri:2017hat} is used to form individual particle tracks 
	from the measured ionization signals.
	Particle momenta are determined from the measured track length for protons and multiple Coulomb scattering pattern for muons~\cite{Abratenko:2017nki}.

	The analysis presented here is performed on data collected from the BNB beam, with an exposure of $4.59 \times 10^{19}$ protons on target (POT). At nominal running conditions, one neutrino interaction is expected in approximately 500 BNB beam spills. A trigger based on scintillation light detected by 32 photomultiplier tubes (PMTs) increases the fraction of recorded spills with a neutrino interaction to $\approx 10 \%$. Application of additional software selection further rejects background events, mostly from cosmic muons, to provide a sample that contains a neutrino interaction in $\approx 15 \%$ of selected spills~\cite{Kaleko:2013eda,Adams:2018gbi}.  CCQE-like event selection, further cosmic rejection and neutrino-induced background rejection are described in Ref.~\cite{Adams:2018lzd}. Muon-proton pair candidates are identified by requiring two tracks with a common vertex and an energy deposition profile consistent with a proton and a muon~\cite{Adams:2016smi}. Further cuts on the track pair opening angle ($|\Delta \theta_{\mu,p} - 90^\circ| < 55^\circ$) and the muon and proton track lengths ($l_\mu > l_p$) reduce the cosmic background rate to less than $1 \%$~\cite{Adams:2018lzd}.


The selected CC1p0$\pi$ event definition includes events with any number of 
protons with momenta below 300 MeV/$c$, neutrons at any momenta, and charged pions with momentum lower than 70 MeV/$c$.
	The minimal proton momentum requirement of 300 MeV/$c$  is guided by its stopping range in LAr and corresponds to five wire pitches in the TPC, to ensure  an efficient particle identification.
	
	To avoid contributions from cosmic tracks, our CC1p0$\pi$ selection considers only pairs of tracks with a fully-contained proton candidate 
	and a fully or partially contained muon candidate in the fiducial volume 
	of the MicroBooNE detector. 
	The fiducial volume is defined by 3  $\textless\,x\,\textless$ 253 cm, -110  $\textless\,y\,\textless$ 110  cm, and 5 $\textless\,z\,\textless$  1031 cm.
The  $x$ axis points  along  the  negative  drift  direction  with  $0$ cm  placed  at  the  anode  plane,  $y$ points vertically upward with $0$ cm at the center of the detector, and $z$ points  along  the  direction  of  the  beam,  with  $0$ cm  at  the  upstream edge of the detector.
Tracks are fully contained if both the start point and end point are within this volume and partially contained if only the start point is within this volume.


We limit our analysis to a phase space region where the detector response  to our signal is well understood
and its effective detection efficiency is higher than 2.5\%.
This corresponds to $0.1 < p_{\mu} < 1.5$  GeV/$c$, $0.3 < p_{p} < 1.0$ GeV/$c$,
$-0.65 < \cos\theta_{\mu} < 0.95$, and $\cos\theta_{p} >  0.15$.
Additional kinematical selections are used to enhance the contribution of CCQE interactions in our \CCIpOpi\ sample. These include requiring 
that the measured muon-proton pairs be coplanar ($|\Delta \phi_{\mu,p} - 180^\circ| < 35^\circ$) relative to the beam axis, 
have small missing transverse momentum relative to the beam direction ($p_T = |\vec{p}^{\,\mu}_T + \vec{p}^{\,p}_T| < 350$ MeV/$c$),
and have a small energy deposition around the interaction vertex that is not associated with the muon or proton tracks.

After the application of the event selection requirement, we retain 410 \CCIpOpi\ candidate events.
We estimate that our \CCIpOpi\ CCQE--like event selection purity equals $\approx$ 84\%~\cite{Adams:2018lzd},
with 81\% of the measured events originating from an underlying CCQE interaction as defined by the GENIE event generator.
The efficiency for detecting \CCIpOpi\ CCQE--like events, out of all generated \CCIpOpi\ with an interaction vertex within our fiducial volume, was estimated using our Monte Carlo (MC) simulation and equals $\approx$ 20\%~\cite{Adams:2018lzd}.
We note that this efficiency includes acceptance effects, as the typical LArTPC efficiency for reconstructing a contained high-momentum proton or muon track is grater than $\sim 90\%$~\cite{Acciarri:2017hat}.

We report single differential cross sections in measured proton and muon kinematics. 
The differential cross section is given by:
\begin{equation}
\label{eq:Xsec}
	\frac{\mathrm{d}\sigma}
	{\mathrm{d}X_n }
	=
	\frac{N^\textrm{on}_n - N^\textrm{off}_n - B_{n}}{ \epsilon_{n} \cdot \Phi_\nu \cdot N_{\textrm{target}} \cdot \Delta^p_{n}},
\end{equation}
where, $X = p_\mu,  \cos\theta_\mu, \phi_\mu,  p_p,  \cos\theta_p, \phi_p$ stands for the kinematical variable that the cross section is differential in and $n$ marks the cross-section bin.
In each bin $n$, $N_n^\textrm{on}$ is the number of measured events when the beam is on,
$N_n^\textrm{off}$ is the number of measured events when the beam is off (i.e., cosmic-induced background events),
$B_n$ is the beam-related background (estimated from MC simulation), $N_{\textrm{target}}$ is the number of scattering nuclei,
$\Phi_\nu$ is the integrated incoming neutrino flux,
$\Delta^\mu_{n}$ and $\Delta^p_{n}$ are the differential bin widths, 
and $\epsilon_{n}$ is the effective particle detection efficiency.

As the detection efficiency is a multidimensional function of the interaction vertex and the particle momentum and direction,
the data were binned in three-dimensional momentum, in-plane, and out-of-place angle bins 
with the effective detection efficiency calculated for each such bin separately and integrated over the interaction vertex in the detector.
The efficiency was extracted based on simulation and is defined as the ratio of the number of reconstructed CC1p0$\pi$ events to the 
number of true generated CC1p0$\pi$ events (with a vertex inside our fiducial volume) in bin n.  
This procedure accounts for bin migration effects such that cross-sections are obtained as a function of real (as oppose to experimentally reconstructed) kinematical variables. 
The results presented herein include the bin migration corrections, which generally have a small impact on the nominal cross-section values as compared with the total cross-section uncertainties (see supplementary materials).
The proton and muon efficiencies were extracted independently of each other (rather than from a full six--fold binning),
such that when the cross-section is differential in muon kinematics the proton kinematics is integrated over and vise-versa.
This is done due to the limited data and simulation statistics and is justified since the proton and muon efficiencies are largely independent in the region of interest. The effect of residual correlations is accounted for in the systematic uncertainties.
We further note that the missing transverse momentum requirement increases the sensitivity of our efficiency corrections to the meson exchange current (MEC) and final state interaction (FSI) models used in our simulations. We accounted for the model sensitivity in our systematic studies detailed below.

%

\begin{table}[b]
\caption{ Integrated cross section values and $\chi^2$ values for the agreement between the measured  cross sections and various event generators.
Results are listed for the full measured phase space and for a limited one of $\cos(\theta_{\mu}) < 0.8$.}
\resizebox{0.47\textwidth}{!}{%
\begin{tabular}{|c|c|c|c|}
\hline
\multicolumn{2}{|c|}{\multirow{2}{*}{}} & \multicolumn{2}{c|}{Integrated Cross Section $[10^{-38} $cm$^2]$}  \\ 
\multicolumn{2}{|c|}{\multirow{2}{*}{}} & \multicolumn{2}{c|}{(Differential Cross Section  $\chi^2$/d.o.f)}  \\ \cline{3-4} 
\multicolumn{2}{|c|}{}                  & $-0.65<\cos(\theta_\mu)<0.95$   & $-0.65<\cos(\theta_\mu)<0.8$                   \\ \hline
\multicolumn{2}{|c|}{Data CC$1p0\pi$ Integrated}       & 4.93 $\pm$ 1.55           & 4.05 $\pm$ 1.40                                                                                                                      \\ \hline 
\multirow{4}{*}{\rotatebox{90}{Generators\;\;\;\;\;}}  & GENIE Nominal           & 6.18~(63.2/28)                      & 4.04~(30.1/27)                      \\ 
                                             & GENIE v3.0.6            & 5.45~(34.6/28)                      & 3.66~(21.4/27)                        \\
                                             & NuWro 19.02.1           & 6.67~(76.7/28)                      & 4.39~(29.9/27)                    \\ 
                                             & NEUT v5.4.0             & 6.64~(78.5/28)                      & 4.39~(32.2/27)                    \\ 
                                             & GiBUU 2019              & 7.00~(82.2./28)                      & 4.78~(40.0/27)                     \\ \hline
\end{tabular}}
\label{IntegratedXSec}
\end{table}


\begin{figure}[t]
	\centering  
        \includegraphics[width=0.75\linewidth]{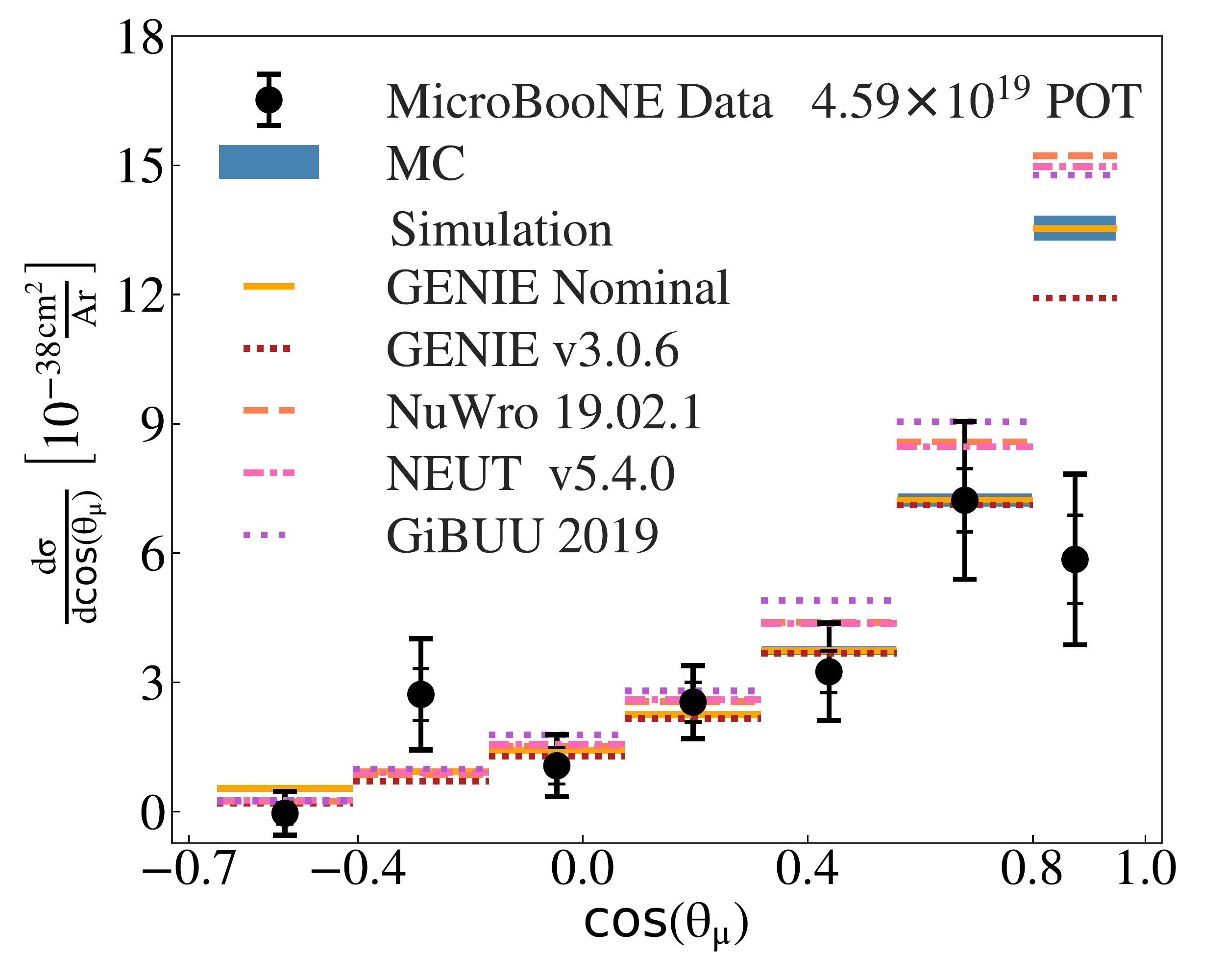}\\
	\caption{The flux integrated single differential \CCIpOpi\ cross sections as a function of the cosine of the measured muon scattering angle.
	Inner and outer error bars show the statistical and total (statistical and systematic) uncertainty at the 1$\sigma$, or 68\%, confidence level. Colored lines show the results of 
	theoretical absolute cross section calculations using different event generators (without passing through a detector simulation). 
	The blue band shows the extracted cross section obtained 
	from analyzing MC events propagated through our full detector simulation. The width of the band denotes the simulation statistical uncertainty.}
	\label{fig:Xsec_1D}
\end{figure}

	    \begin{figure*}[t]
		\centering  
		\includegraphics[width=0.25\linewidth]{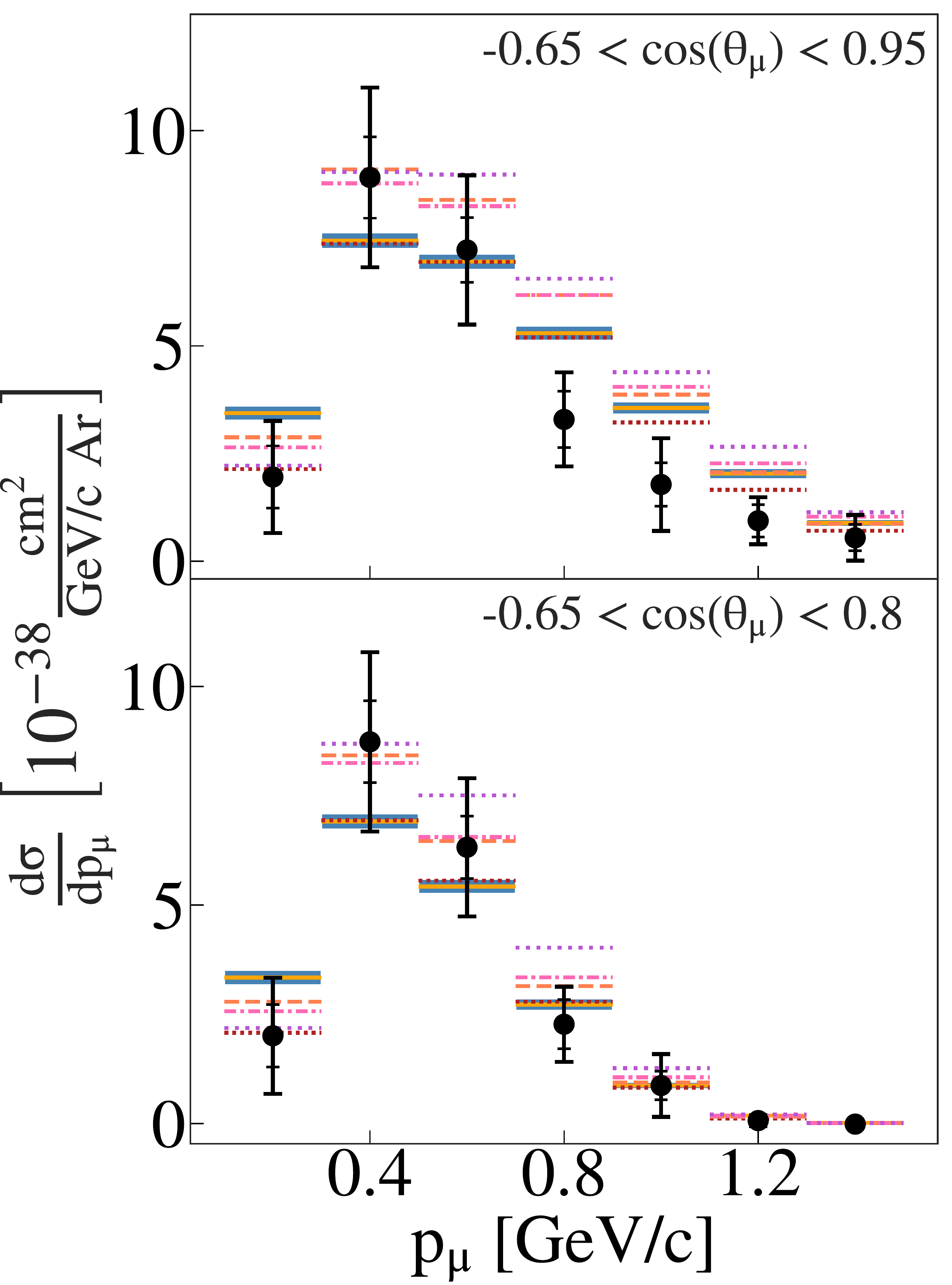}
		\includegraphics[width=0.25\linewidth]{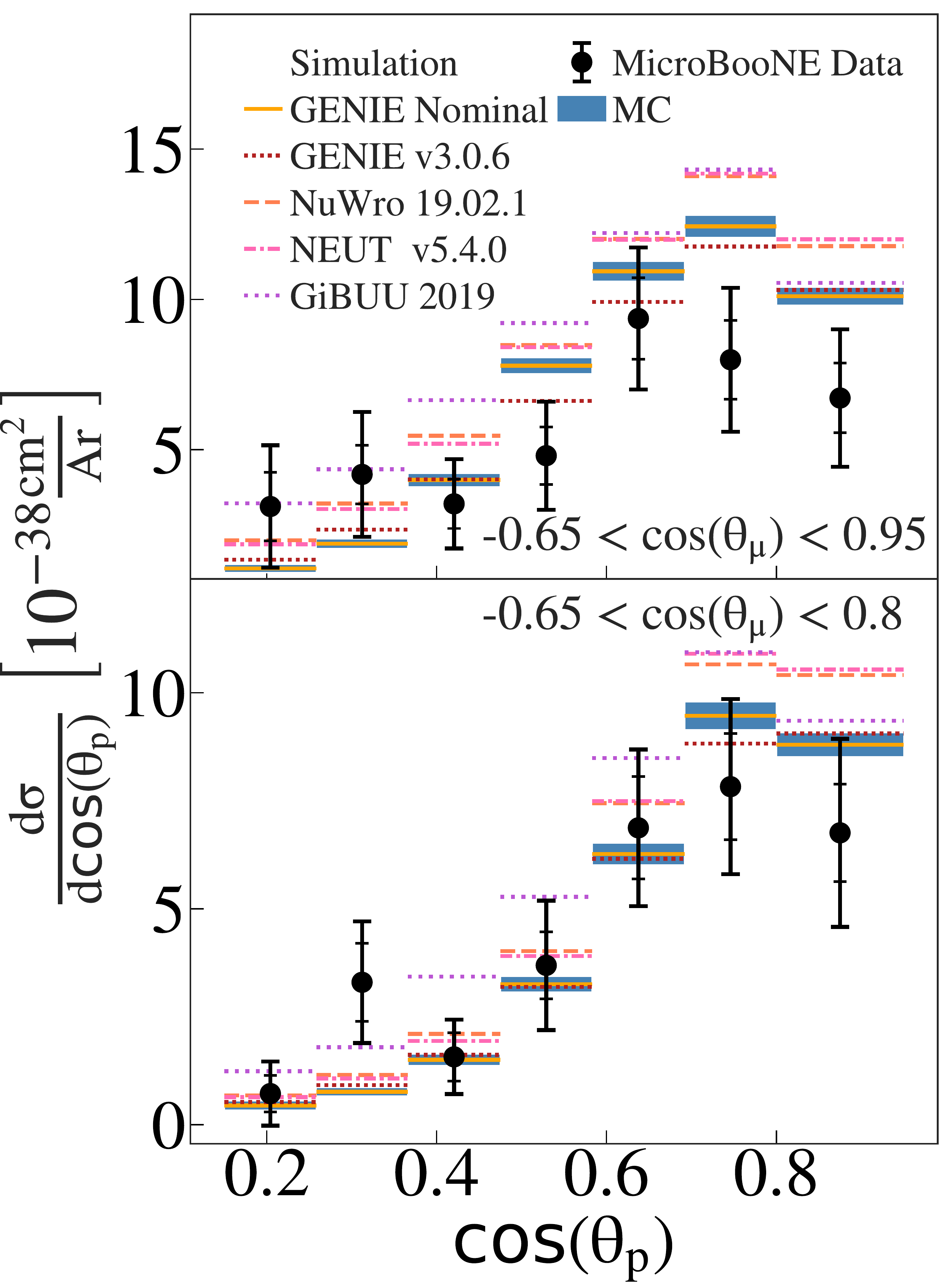}		
		\includegraphics[width=0.25\linewidth]{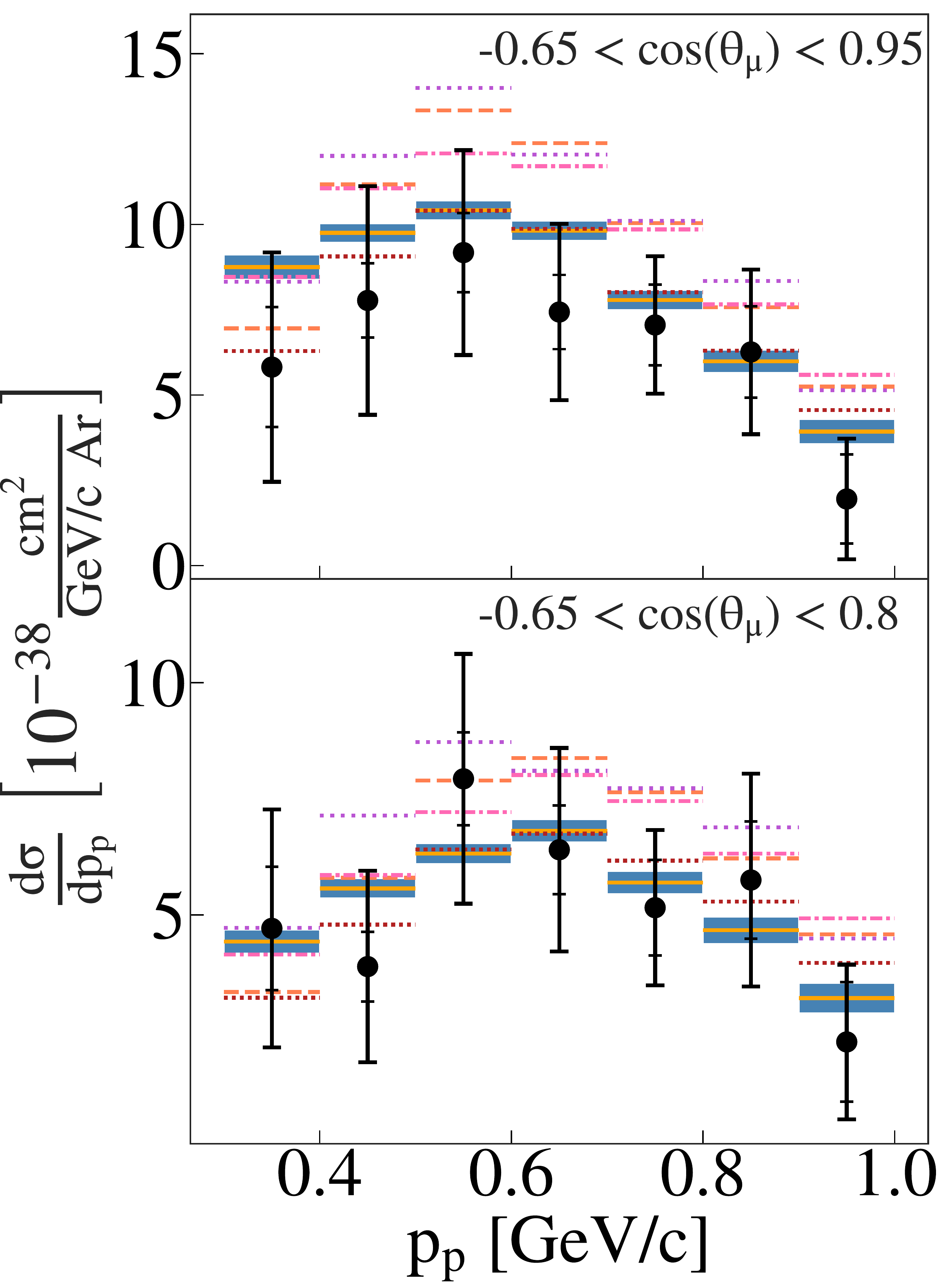}
		\caption{As Fig.~\ref{fig:Xsec_1D},  but for the differential cross sections as a function of measured muon momentum (left) and measured proton scattering angle (middle) 
		and momentum (right). 
		Cross sections are shown for the full measured phase-space (top) and for events with cos$(\theta_\mu) < 0.8$ (bottom).}
		\label{fig:Xsec_1D_without_last_bin}
	\end{figure*}

The extracted cross sections are expected to be independent of the azimuthal angle $\phi$. 
However, the simple model used to simulate the effect of induced charge on neighboring TPC wires leads to a low reconstruction efficiency of tracks perpendicular to the wire planes 
($\phi \approx 0$ and $\phi \approx \pm\pi$) that created an artificial $\phi$ dependence to the cross section. 
We correct for this effect using an iterative procedure. We first reweight events with a muon track falling in the $\phi \approx 0$ bin and $|\sin\theta|>0.3$ to the weighted 
average of the cross sections in all other bins of $\phi_{\mu}$ where  $|\sin\theta|>0.3$. Due to the coplanarity requirement, this reweighting affects the 
distribution of $\phi_{p} \approx \pm \pi$. We repeat the process starting from a proton track with $\phi_{p} \approx 0$ until the cross section change is less than 0.01$\%$, 
typically after 5 iterations.

The integrated measured CC1p0$\pi$ cross section is summarized in Table~\ref{IntegratedXSec}. 
The statistical uncertainty of our measurement is 15.9\%.
The systematic uncertainty sums to 26.2\%
and includes contributions from the neutrino flux prediction and POT estimation (18.7\%),
detector response modeling (18.4\%),
imperfect proton and muon efficiency decoupling (5.7\%), 
and neutrino interaction cross section modeling (7.1\%).

The neutrino flux is predicted using the flux simulation of the MiniBooNE Collaboration that used the same beam line~\cite{Aguilar-Arevalo:2013dva}. We account for the small distance 
between 
MiniBooNE and \uB.
Neutrino  cross  section  modeling uncertainties were estimated using the GENIE framework of event reweighting \cite{Andreopoulos:2009rq,Andreopoulos:2015wxa} with its standard reweighting 
parameters.  
For both cross section and 
flux systematics, we use a multisim technique \cite{Roe:2007hw},  which  consists  of  generating many MC replicas, each one called a ``universe''\kern-.2em, where model parameters are varied 
within their uncertainties. Each universe represents a different reweighting. The simultaneous reweighting of all model parameters allows the correct treatment of their correlations.

A different model is followed for detector model systematic uncertainties, 
that are dominated by individual detector parameters.
Unisim samples \cite{Roe:2007hw} are generated, where one detector parameter is varied each time by $1\sigma$.  We then examine the impact of each parameter variation on the extracted 
cross sections, by obtaining the differences with respect to the central value on a bin--by--bin basis.
We note that the detection efficiency used for the cross section extraction is re-evaluated for each variation separately, including bin migration corrections. This procedure therefore accounts for the systematic uncertainty in these corrections due to both the cross-section and detector response modeling.
One exception to this process is the systematic uncertainty due to induced charge effects mentioned above that include the data-driven correction and are thus estimated separately (see supplementary materials).
We then define the total detector $1\sigma$ systematic uncertainty by summing in quadrature the effect of each individual variation. 

A dedicated MC simulation was used to estimate possible background from events in which a neutrino interacts outside the \uB\ cryostat but produce particles that enter the TPC and pass the event selection cuts~\cite{Abratenko:2019jqo}. No such events were found in that study, which is also supported by our observation that the z-vertex distributions for the measured events follows a uniform distribution (see supplementary materials).

The MC simulation used to estimate the backgrounds and effective efficiency contains real cosmic data overlayed onto a neutrino interaction simulation that uses 
GENIE~\cite{Andreopoulos:2009rq,Andreopoulos:2015wxa} to simulate both the signal events and the beam backgrounds.
See Ref.~\cite{Adams:2018lzd} for details. For the simulated portion, the particle propagation is based on GEANT4~\cite{Geant4}, while the simulation of the MicroBooNE detector 
is performed in the LArSoft framework~\cite{Pordes:2016ycs,Snider:2017wjd}. 
The beam--related background subtracted from the \CCIpOpi\ events is simulated.

Fig.~\ref{fig:Xsec_1D} shows the flux integrated single differential \CCIpOpi\ cross section 
as a function of the cosine of the measured muon scattering angle.
The data are compared to several theoretical 
calculations and to our GENIE-based MC prediction.
The latter is the result of analyzing a sample of MC events produced using 
our ``nominal'' GENIE model and propagated through the full detector simulation in the same way as data. 

This model (GENIE v2.12.2)~\cite{Andreopoulos:2009rq,Andreopoulos:2015wxa} treats the nucleus as a  
the Bodek-Ritchie Fermi Gas, used
the Llewellyn-Smith CCQE scattering prescription~\cite{LlewellynSmith:1971uhs},
and the empirical MEC model~\cite{Katori:2013eoa},
 Rein-Sehgal resonance (RES) and coherent scattering (COH) model~\cite{Rein:1980wg},
 a data driven FSI model denoted as \enquote{hA}~\cite{Mashnik:2005ay}.

In addition, theoretical predictions by several other event generators are shown at the cross-section level (i.e with no detector simulations)~\cite{Stowell_2017}.
These include GENIE v2.12.2 and v3.0.6~\cite{Andreopoulos:2009rq,Andreopoulos:2015wxa}, NuWro 19.02.1~\cite{GolanNuWro:2008yp}, and NEUT v5.4.0~\cite{Hayato:2008yp}  (see supplementary materials).
The agreement between the ``nominal'' GENIE calculation (v2.12.2) and the MC prediction constitutes a closure test for our analysis. 
The other generators all improve on GENIE v2.12.2 by using updated nuclear interaction models, 
among which is the use of a Local Fermi Gas model~\cite{Carrasco:1989vq} and
Random Phase Approximation correction~\cite{RPA}.
GENIE v3.0.6 also includes Coulomb corrections for the outgoing muon~\cite{Engel:1997fy}.
The theoretical models implemented in these event generators include free parameters that are typically fit to data,
with different generators using different data sets.
We also consider the GiBUU 2019~\cite{Mosel:2008yp} event generator which fundamentally differs from the others due to its use of a transport equation approach.

As can be seen in Fig.~\ref{fig:Xsec_1D}, all models are in overall good agreement with our data, except for the highest \CosThetaMu\ bin, where the measured cross 
section is significantly lower than the theoretical predictions.
This discrepancy cannot be explained 
by the systematic uncertainties and is therefore indicative of an issue with the theoretical models. 
Specifically, high \CosThetaMu\ correspond to low momentum transfer events which were previously observed to not be well reproduced by theory in inclusive reactions~\cite{Abratenko:2019jqo,Carneiro:2019jds} and is now also seen in exclusive reactions.
We note that the high \CosThetaMu\ bin has large beam-related background ($B_n$ in Eq.~\ref{eq:Xsec}), that is estimated using the GENIE v2.12.2 based MC simulation (see supplementary materials).

As the differential cross sections in proton kinematics and muon momentum include contributions from all muon scattering angles, their agreement with the theoretical calculation is affected by this disagreement. Fig.~\ref{fig:Xsec_1D_without_last_bin} shows this comparison between the relevant cross sections in the full available phase-space (top) and in the case where events with \CosThetaMu\ $ > 0.8$ are excluded (bottom).  Removing this part of the phase-space significantly improves the agreement between data and theory. 

Table~\ref{IntegratedXSec} also lists the $\chi^2$  for the agreement of the different models with the data for differential cross sections for the full available phase-space and for \CosThetaMu\ $ < 0.8$. Systematic uncertainties and correlations were accounted for using covariance matrices.  The $\chi^2$ values reported in the table are the simple sum of those $\chi^2$ values obtained for each distribution separately.  
As can be seen GENIE v3.0.6 is the only model that reaches a $\chi^2$/d.o.f. close to unity for the full phase-space.
It is also the closest model to the data at the highest \CosThetaMu\ bin.
For all other models, the $\chi^2$/d.o.f. in the \CosThetaMu\ $ < 0.8$ sample is reduced by a factor of $\sim 2$ as compared to the full phase-space sample. 
GENIE v3.0.6 shows a smaller reduction in this case, 
and GiBUU 2019 obtains a consistently higher $\chi^2$/d.o.f. for both the full and limited phase-space samples.

The improved agreement with the data observed for GENIE v3.0.6, especially for the full phase-space sample, is intriguing. 
Specifically, GENIE v3.0.6 and NEUT v5.4.0 are quite similar, using the same nuclear, QE, and MEC models, 
which are the most significant processes in our energy range.
They do differ in the coulomb corrections that only GENIE v3.0.6 has, their free parameter tuning process, 
and the implementation of RPA correction, that are known to be important at low momentum transfer~\cite{RPA}.
Our data indicates that these seemingly small differences can have a highly significant impact, as seen in table~\ref{IntegratedXSec}.

    \begin{figure}[t]
		\centering  
 		\includegraphics[width=0.48\linewidth]{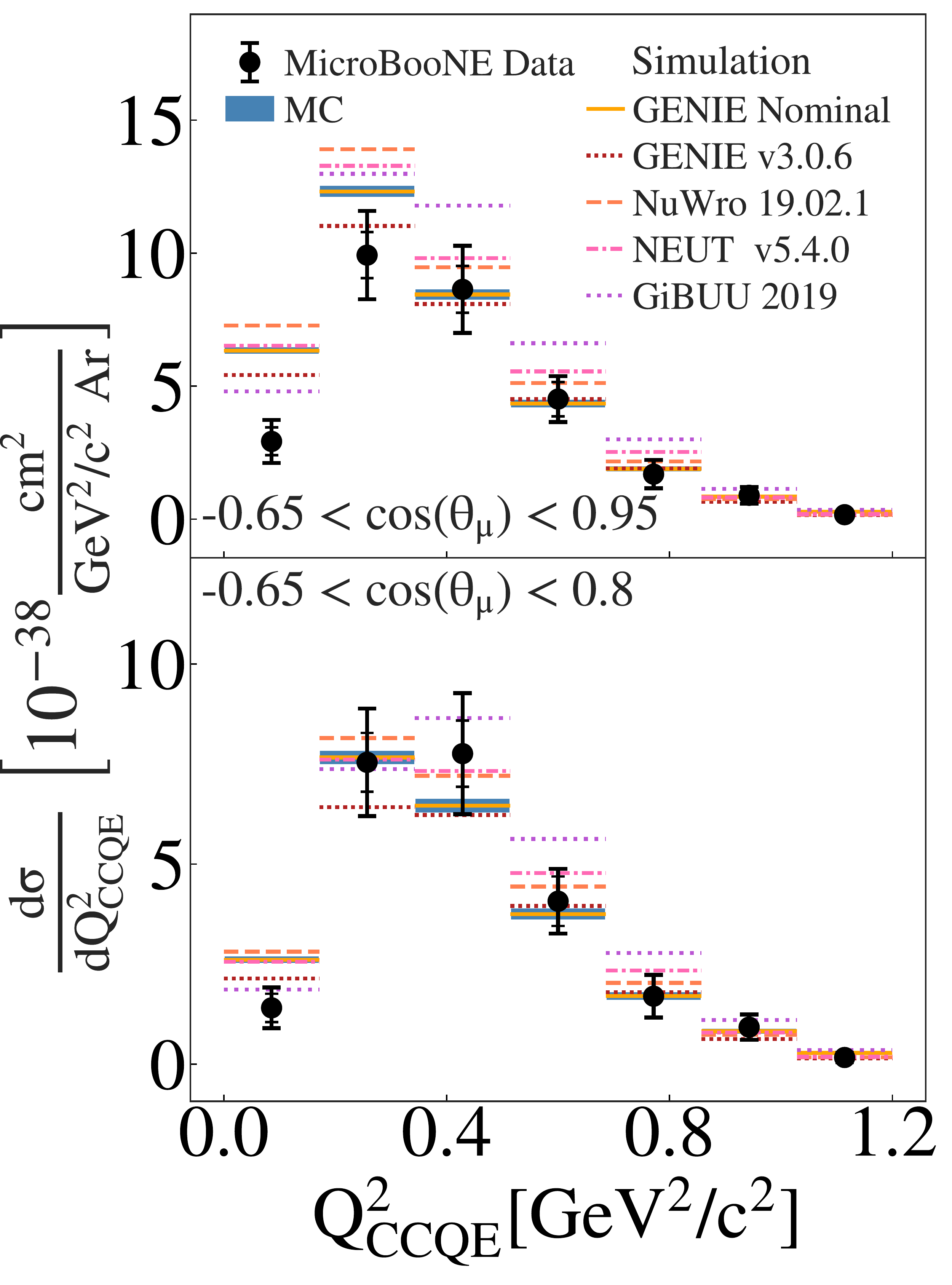}
 		\includegraphics[width=0.48\linewidth]{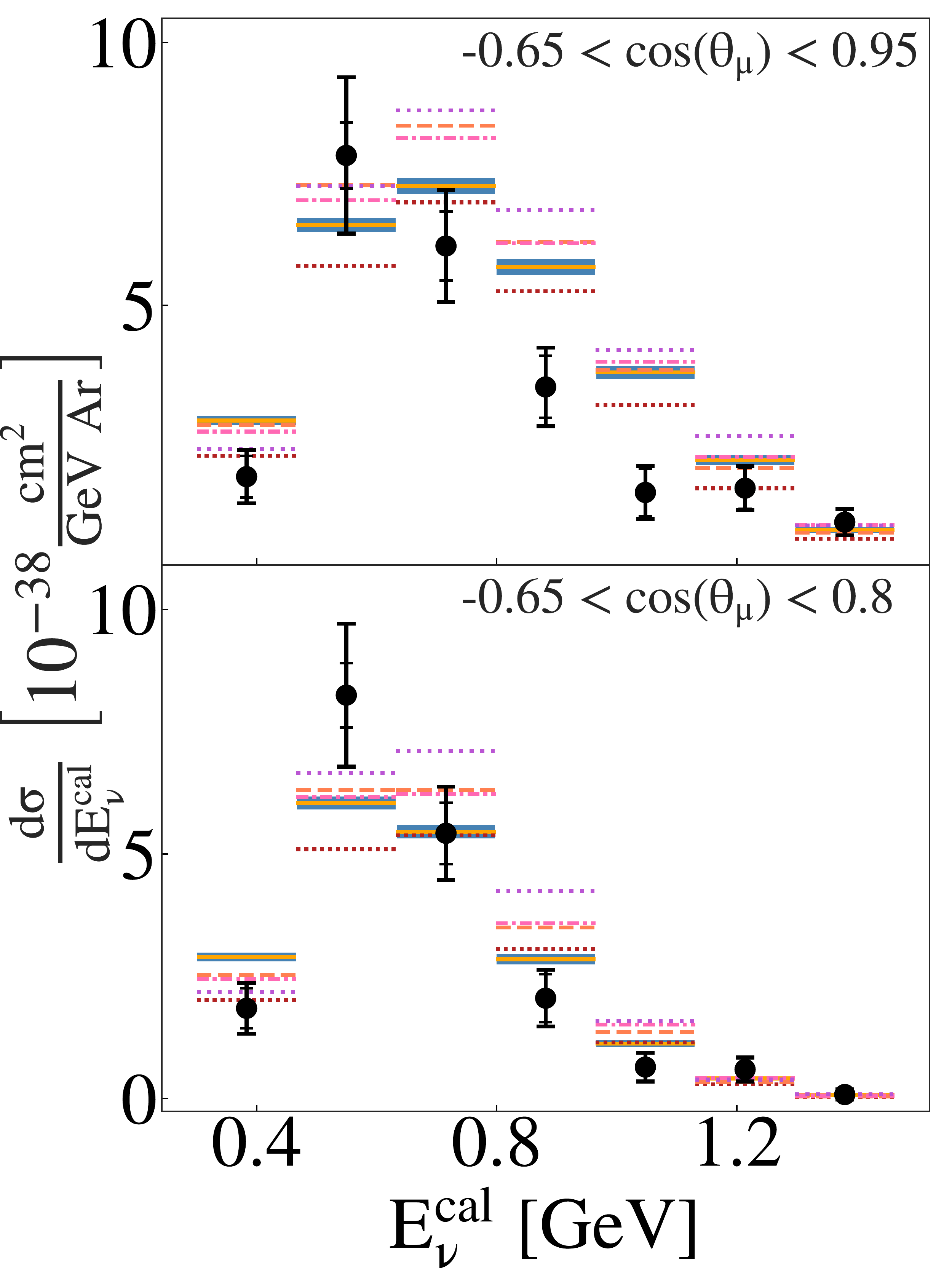} 		
		\caption{The flux integrated single differential \CCIpOpi\ cross sections as a function of 
		$Q^{2}_{CCQE}  = (E^{cal}_{\nu} - E_{\mu})^{2} - (\vec{p}_{\nu} - \vec{p}_{\mu})^{2}$ and $E_{\nu}^{cal}  = E_{\mu} + T_{p} + BE$, 
		where $BE =  40$ MeV and $\vec{p}_{\nu} = (0, 0, E^{cal}_{\nu})$. 
		Inner and outer error bars show the statistical and total (statistical and systematic) uncertainty at the 1$\sigma$, or 68\%, confidence level. 
		Colored lines show the results of theoretical absolute cross section calculations using different event generators (without passing through a detector simulation). 
		The blue band shows the extracted cross section obtained from analyzing MC events passed through our full detector simulation.}
		\label{fig:Xsec_1D_Ev_Q2}
	\end{figure}

Lastly, Fig.~\ref{fig:Xsec_1D_Ev_Q2} shows the flux-integrated single differential cross sections as a function of calorimetric measured energy and reconstructed momentum transfer, 
with and 
without events with \CosThetaMu\ $> 0.8$. The former is defined as $E_{\nu}^{cal} = E_{\mu} + T_{p} + BE$, and the latter as 
$Q^{2}_{CCQE} = (\vec{p}_{\nu} - \vec{p}_{\mu})^{2} - (E^{cal}_{\nu} - E_{\mu})^{2}$, where E$_{\mu}$ is the muon energy, T$_{p}$ is the proton kinetic energy, BE = 40 MeV 
is the effective 
nucleon binding energy for \argon{}, and $\vec{p}_{\nu} = (0, 0, E^{cal}_{\nu})$ is the reconstructed interacting neutrino momentum. 
$E_{\nu}^{cal}$ is often used as a proxy for the reconstructed neutrino energy.  

Overall, good agreement is observed between data and calculations for these complex variables, even for the full event sample without the   \CosThetaMu\ $ < 0.8$  requirement.

In summary, 
we report the first measurement of $\nu_\mu$ CCQE--like differential cross sections on \argon\ 
for event topologies with a single muon and a single proton detected in the final state.
The data are in good agreement with GENIE predictions, except at small 
muon scattering angles that correspond to low momentum-transfer reactions.
This measurement confirms and constrains calculations essential for the extraction of oscillation 
parameters and highlights kinematic regimes where improvement of theoretical models is required.
The benchmarking of exclusive \CCIpOpi\ cross sections on \argon presented here suggests that
measurements of \CCIpOpi\ interactions are a suitable choice for use in precision neutrino 
oscillation analyses, especially after theoretical models are reconciled with the small scattering angle data.

\begin{acknowledgments}
This document was prepared by the MicroBooNE collaboration using the resources of the Fermi National Accelerator Laboratory (Fermilab), a U.S. Department of Energy, Office of Science, HEP User Facility. Fermilab is managed by Fermi Research Alliance, LLC (FRA), acting under Contract No. DE-AC02-07CH11359.  MicroBooNE is supported by the following: the U.S. Department of Energy, Office of Science, Offices of High Energy Physics and Nuclear Physics; the U.S. National Science Foundation; the Swiss National Science Foundation; the Science and Technology Facilities Council (STFC), part of the United Kingdom Research and Innovation; and The Royal Society (United Kingdom). Additional support for the laser calibration system and cosmic ray tagger was provided by the Albert Einstein Center for Fundamental Physics, Bern, Switzerland. The work presented in this manuscript was supported in part by the Azrieli Foundation, Israel Science Foundation, Visiting Scholars Award Program of the Universities Research Association, and the Zuckerman STEM Leadership Program.
\end{acknowledgments}

\nocite{suppl,Heck:1998vt,Nieves:2012yz,Schwehr:2016pvn,Nowak:2009se,Kuzmin:2003ji,Berger:2007rq,Graczyk:2007bc,Berger:2008xs,Ashery:1981tq,Bodek:2017hat,Graczyk:2007bc,Leitner:2006ww,Mosel:2019vhx,Sjostrand:2006za}.

	\bibliography{../CCQE_bib}
	
\end{document}